\documentclass[journal]{IEEEtran}

\usepackage{amssymb}
\usepackage{amsmath}
\usepackage{bm}
\usepackage{amsthm}
\usepackage{graphicx}
\usepackage{subfigure}
\usepackage{float}
\usepackage{cite}
\usepackage{enumerate}
\usepackage{enumitem}
\usepackage{multirow}
\usepackage{amsfonts}
\usepackage{makecell} 
\usepackage{url}
\usepackage{amsthm}
\usepackage{color, soul}

\begin{document}
	
\title{\huge{Intelligent Omni-Surfaces for Full-Dimensional Wireless Communications: Principle, Technology, and Implementation}}
\author{
{Hongliang Zhang}, 
{Shuhao Zeng}, 
{Boya Di}, 
{Yunhua Tan}, 
{Marco Di Renzo}, \\
{M\'erouane Debbah}, 
{Zhu Han}, 
{H. Vincent Poor},
{and Lingyang Song}

\thanks{This work has been submitted to the IEEE for possible publication. Copyright may be transferred without notice, after which this version may no longer be accessible.}

\thanks{H. Zhang is with Department of Electronics, Peking University, Beijing, China, and also with Department of Electrical and Computer Engineering, Princeton University, NJ, USA.}

\thanks{S. Zeng, B. Di, Y. Tan, and L. Song are with Department of Electronics, Peking University, Beijing, China.}

\thanks{M. Di Renzo is with Universit\'e Paris-Saclay, CNRS, CentraleSup\'elec, Laboratoire des Signaux et Systemes, Gif-sur-Yvette, France.}

\thanks{M. Debbah is with Universit\'e Paris-Saclay, CNRS, CentraleSup\'elec, Gif-sur-Yvette, France and the Lagrange Mathematical and Computing Research Center, Paris, France.}

\thanks{Z. Han is with Electrical and Computer Engineering Department, University of Houston, Houston, TX, USA, and also with the Department of Computer Science and Engineering, Kyung Hee University, Seoul, South Korea.}

\thanks{H. V. Poor is with Department of Electrical and Computer Engineering, Princeton University, NJ, USA.}
\vspace{-6mm}}

\maketitle

\begin{abstract}
The recent development of metasurfaces has motivated their potential use for improving the performance of wireless communication networks by manipulating the propagation environment through nearly-passive sub-wavelength scattering elements arranged on a surface. However, most studies of this technology focus on reflective metasurfaces, i.e., the surface reflects the incident signals towards receivers located on the same side of the transmitter, which restricts the coverage to one side of the surface. In this article, we introduce the concept of intelligent omni-surface (IOS), which is able to serve mobile users on both sides of the surface to achieve full-dimensional communications by jointly engineering its reflective and refractive properties. The working principle of the IOS is introduced and a novel hybrid beamforming scheme is proposed for IOS-based wireless communications. Moreover, we present a prototype of IOS-based wireless communications and report experimental results. Furthermore, potential applications of the IOS to wireless communications together with relevant research challenges are discussed.
\end{abstract}
\vspace{-3mm}
\section{Introduction}
The past decade has witnessed an explosive growth in the number of mobile devices and wireless data traffic, triggering the urgent needs for innovative communication paradigms \cite{MAMMCJS-2020}. Various technologies have been developed to provide high-speed and seamless data services by exploiting the randomness of the propagation environment, such as relaying and massive multiple input and multiple output (MIMO) systems. However, these techniques often require complex hardware implementations with an inevitable increase of the power consumption and a higher signal processing complexity \cite{HBLZ-2021}. Besides, these technologies can only adapt to the propagation environment but cannot control it, which makes the quality-of-service~(QoS) in harsh propagation environments non-guaranteed \cite{MKJ-2020}.

Fortunately, the recent development of metasurfaces has given rise to the possibility of controlling the electromagnetic waves in profoundly new ways, and has motivated the exploration of new smart surface technologies for wireless applications \cite{MHLKZG-2020}. A metasurface is an ultra-thin engineered surface, containing multiple nearly-passive sub-wavelength scattering elements, which enables exotic manipulations of the signals impinging upon it. These signal transformations can be realized through an appropriate configuration of, e.g., positive intrinsic negative (PIN) diodes distributed throughout the surface. Depending on the ON/OFF state of each PIN diode, the amplitude and phase  of the scattered signals can be adjusted~\cite{CBWJXX-2017}. These programmable characteristics of the metasurfaces enable the shaping of the propagation environment as desired, and allow the re-transmission of signals to the receiver at a reduced cost, size, weight, and power~\cite{MMDAMCVGJHJAGM-2019}. 

In the wireless communications literature, a widely studied implementation of metasurfaces is referred to as intelligent reflecting surfaces (IRSs), which consists of reflecting the signals from one side of the surface towards users located on the same side of it~\cite{BHLYZH-2020}. However, this implies that the users located on the opposite side of the metasurface are out of the IRS's coverage. To address this issue, we introduce an innovative implementation of metasurfaces, which we refer to as intelligent omni-surface (IOS). In contrast to an IRS, the proposed IOS has the dual functionality of signal reflection and refraction \cite{DOCOMO}. To be specific, the IOS can simultaneously reflect and refract the signals that impinge on one side of the surface towards mobile users (MUs) located on both sides, respectively. The power ratio of the reflected and refracted signals can be optimized and is determined by the structure of each IOS element. Equipped with the capability of joint reflection and refraction, an IOS can achieve full-dimensional communications with the MUs wherever their locations are with respect to the IOS \cite{SHBYZL}. As a result, the QoS of all communication links in the network can be improved.

In this article, we introduce the working principle of the IOS and we study the feasibility of utilizing the IOS for realizing full-dimensional communications. More precisely, we provide the following contributions.
\begin{itemize}
    \item \emph{Working Principle:} Unlike an IRS, the proposed IOS can split the energy into reflected and refracted signals with different amplitudes and phases.
    
    \item \emph{Hybrid Beamforming Scheme:} To serve MUs on both sides of the IOS, we propose a hybrid beamforming scheme for IOS-based wireless communications~\cite{SHBYMZHL}. In particular, digital beamforming is performed at the base station (BS) and analog beamforming is implemented at the IOS.
    
    \item \emph{Prototype and Evaluation:} To substantiate the feasibility of full-dimensional communications, we build an IOS-based wireless communication prototype and present experimental results that prove that an IOS can effectively control angles of the reflected and refracted beams. These experimental results complement the recent design of beamforming algorithms for IOS-aided communications \cite{SHBYMZHL}, and mark the novel contribution of this article. 
\end{itemize}

Broadly speaking, an IOS can be deployed anywhere it is appropriate or necessary to control the propagation of the radio waves for enhancing the communication performance. However, as the configuration of the IOS requires accurate channel estimation algorithms, which is challenging in IOSs that comprise a large number of scattering elements, an IOS is expected to be more easily deployed in scenarios that are characterized by a low mobility and a low channel variability. Potential use cases for IOS-aided wireless communications include the following.
\begin{itemize}
    \item \emph{Coverage Extension:} By deploying an IOS at the cell edge, the users within the cell coverage and those outside the cell coverage area can be simultaneously served by capitalizing on the joint reflection and refraction capabilities of the IOS. 
    
    \item \emph{Interference Cancellation:} In ultra-dense networks, some users may be located within the coverage area of multiple small cells, which increases the multi-cell interference. An IOS can help focus the signals towards specified users on both sides of the surface while reducing the interference towards the users located in other cells.
    
    \item \emph{Secure Communications:} With an IOS, users on one side of the IOS can receive signals from users located on the same side of the surface, however, the refracted signals may not be received by the users on the opposite side of the surface. This functionality may be utilized to realize secure communications.
    
    \item \emph{Sensing and Localization:} Sensing and localization through RF signals are important applications in future wireless networks. The capability of the IOS to shape the propagation environment can be leveraged for improving the sensing and localization accuracy.
\end{itemize}

The rest of this article is organized as follows. First, we introduce the working principle of the IOS in Section \ref{sec:principle}. Then, a novel hybrid beamforming design for IOS-based wireless communications is presented in Section \ref{sec:beamforming}. A hardware prototype of the IOS is presented and experimental results are illustrated in Sections \ref{sec:prototype} and \ref{sec:experiment}, respectively. Potential applications of the IOS in wireless communications and related research problems are discussed in Section~\ref{sec:applications}. Finally, conclusions are drawn in Section \ref{sec:conclusion}.

\vspace{-3mm}
\section{IOS: Working Principle}
\label{sec:principle}
\begin{figure}[!t]
	\centering
	\includegraphics[width=0.40\textwidth]{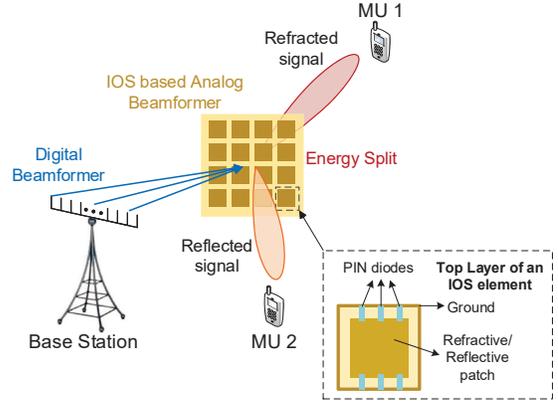}
	\vspace{-3mm}
	\caption{An illustration of the IOS-based hybrid beamforming for two MUs.}
	\vspace{-6mm}
	\label{angle}
\end{figure}

An IOS is a two-dimensional engineered surface that comprises electrically-controllable scattering elements. In particular, the considered IOS is made of several reconfigurable elements of equal size. Each element comprises multiple metallic patches and $N$ PIN diodes that are evenly distributed on a dielectric substrate, as shown in Fig. \ref{angle}. The metallic patches are connected to the ground via the PIN diodes that can be switched between their ON and OFF states according to predetermined bias voltages. The ON/OFF state of these PIN diodes determines the amplitude and phase response applied by the IOS to the incident signals. Each metallic patch can be configured in $2^{N}$ different states, which results in $P \le 2^N$ different amplitudes and phases, depending on the implementation of the IOS, that are usually equally distributed in $[0, 2\pi]$. When a signal impinges, from either sides of the surface, upon one element of the IOS, a fraction of the incident power is reflected and refracted towards the same side and the opposite side of the incident signal, respectively \cite{SHBYMZHL}. This capability enables the IOS to adaptively provide coverage to the users wherever their locations are.

Under the assumption of periodic boundary conditions, complex-valued reflection and refraction coefficients of the generic element of the IOS can be defined. In particular, the reflection and refraction coefficients may depend on the direction of incidence and on the phase shift applied by the IOS element. The amplitude of the reflection and refraction coefficients of the generic IOS element are determined by the size of each element. It is worth mentioning that the proposed IOS is flexibly designed to ensure that the amplitude of the reflected and refracted signals can be different. The power ratio between them is, in particular, a constant determined by the structure and hardware implementation of the IOS element. In addition, the phase shifts induced to the reflected and refracted signals can be different, which is also dependent on the structure and hardware implementation of the IOS element. Finally, the IOS is designed to operate over a predetermined frequency band. Outside the frequency range of design, the response of the IOS is, in general, different from the nominal one and cannot be predicted.

\begin{figure*}
    \centering
    \includegraphics[width=0.68\textwidth]{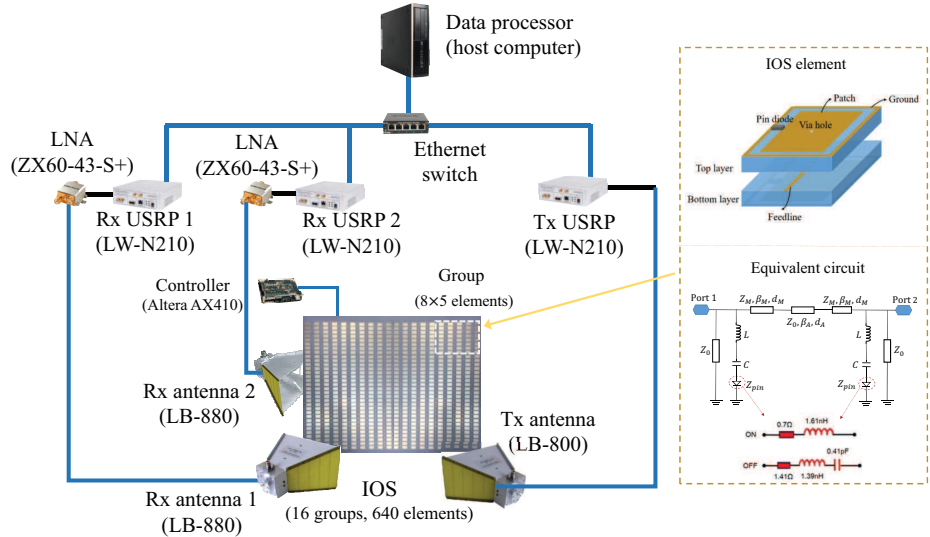}
    \vspace{-3mm}
    \caption{An illustration of the IOS-based wireless communication prototype.}
    \label{prototype}
    \vspace{-6mm}
\end{figure*}

\vspace{-3mm}
\section{IOS-Based Hybrid Beamforming}
\label{sec:beamforming}

To shape the propagation of the radio waves towards specified directions of reflection and refraction, we present a hybrid beamforming scheme for application to the designed IOS. As shown in Fig.~\ref{angle}, \emph{digital beamforming} is performed at the BS and \emph{analog beamforming} is implemented at the IOS. Specially, the BS first encodes multiple data streams via a digital beamformer and transmit them through multiple antennas after up-conversion to a predetermined carrier frequency and after allocating a given transmit power to each transmit antenna. The BS and the IOS exchange information through a wireless control link. Through this wireless link, the BS configures the phase shifts applied by each element of the IOS so as to realize the desired analog beamforming. The IOS splits the incident energy into two parts: some energy is reflected to serve the users located on the same side as the BS, and the rest of the energy is refracted to serve users located on the opposite side. Therefore, an IOS can achieve a full-dimensional coverage. In the following, we elaborate on the hybrid beamforming scheme based on the availability of the channel state information (CSI) at the BS.
\vspace{-2mm}
\subsection{Instantaneous CSI at the BS}
Even assuming that full-instantaneous CSI is available at the BS, it is difficult to compute the optimal digital beamformer at the BS and the configuration of the IOS elements (i.e., the analog beamformer) that optimize the system performance, e.g., the users' sum-rate. The main difficulties include the following:
\begin{itemize}
    \item The digital and the analog beamformers are coupled with each other, which results in a non-convex optimization problem.
    
    \item In practice, the number of states of each IOS element is finite. Therefore, the feasibility set for the optimization of the analog beamformer is discrete, which leads to an NP-hard integer program. 
\end{itemize}

To tackle the first challenge and reducing the computational complexity, the optimization of the digital and analog beamformers can be decoupled. By using some state-of-the-art (but sub-optimal) digital beamforming schemes, the digital beamformer and the states of the IOS can be optimized sequentially. To maximize, for example, the system sum-rate, one can use a zero-forcing digital beamformer, whose closed-form expression is available if the states of the IOS are given. Subsequently, the state of each IOS element can be optimized one by one~\cite{SHBYMZHL}. To tackle the second challenge, some performance bounds can be employed to relax the discrete-valued phase shifts into continuous-valued ones.  

\vspace{-3mm}
\subsection{Statistical CSI at the BS}
A major challenge for the optimization of the digital and IOS analog beamformers lies in the large overhead for acquiring the instantaneous CSI at the BS, since the number of IOS elements may be large. This necessitates the design of optimization algorithms with reduced CSI that decrease the overhead for channel estimation and configuration.

A potential solution to address this issue is to utilize slowly varying statistics of the CSI for optimizing the IOS analog beamformer instead of instantaneous CSI, while relying on instantaneous CSI for the design of the digital beamformer. For example, in a considered period of time, the BS can first transmit some training pilots for obtaining some statistics of the CSI, including the first and second-order moments, which is easier to obtain than the instantaneous CSI. Once the statistical CSI is acquired by the BS, it can optimize the states of the IOS elements accordingly, by using distributionally robust optimization methods \cite{ADM-2020}. Therefore, the states of the IOS elements are optimized based on the statistics of the CSI, which are assumed not to change or to change slowly in the considered coherence time. Once the states of the IOS elements are identified, the BS can dynamically optimize the digital beamformer at each time slot by using instantaneous CSI, since state-of-the-art estimation protocols can be used for this phase. This makes the IOS transparent to currently deployed wireless networks and communication protocols.

\begin{table*}[!t]
\renewcommand\arraystretch{1.2}
\caption{Reflection and Refraction Coefficients of Each IOS Element.}
\centering
\scriptsize
\begin{tabular}{| c | c | c |  p{1.0cm}<{\centering}| p{1.2cm}<{\centering} | p{1.0cm}<{\centering} | p{1.2cm}<{\centering}| p{1.0cm}<{\centering}| p{1.0cm}<{\centering}|}
\Xhline{1.pt}
\multirow{2}*{\textbf{State}} &
\multirow{2}*{\textbf{PIN Diode-1}}  & \multirow{2}*{\textbf{PIN Diode-2}} &
\multicolumn{3}{c|}{\textbf{Reflection Coefficient}} &
\multicolumn{3}{c|}{\textbf{Refraction Coefficient}}\\ 
\cline{4-9} 
	& & & \textbf{Phase} & \textbf{Amplitude} & \textbf{Power} &\textbf{Phase} & \textbf{Amplitude} & \textbf{Power}\\
\hline
$1$ & OFF & OFF & $20^\circ$ & 0.46 & 0.21 & $300^\circ$ & 0.58 & 0.34\\
$2$ & ON & ON & $215^\circ$ & 0.55 & 0.30 & $123^\circ$ & 0.81 & 0.66\\
\Xhline{1.pt}
\end{tabular}
\label{state}
\end{table*}
\begin{figure*}[!t]
	\centering
	\includegraphics[width=0.69\textwidth]{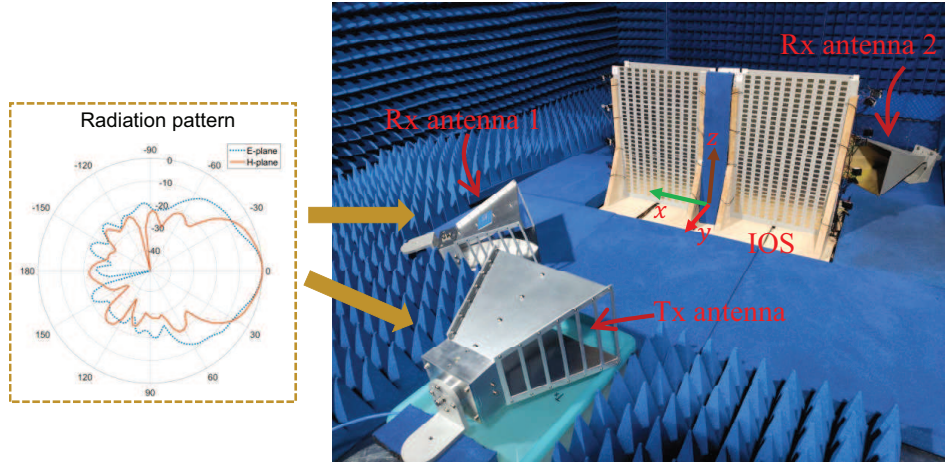}
	\vspace{-3mm}
	\caption{Experimental setup and radiation patterns of antennas.}
	\label{experiment_setup}
	\vspace{-6mm}
\end{figure*}

\vspace{-3mm}
\section{IOS Hardware Prototype}
\label{sec:prototype}

In this section, we describe a hardware prototype to validate the concept of IOS. We first present the implementation of the IOS, and then introduce the specific hardware modules of the prototype.

\vspace{-3mm}
\subsection{Implementation of the IOS}
As shown in Fig.~\ref{prototype}, the designed IOS prototype consists of 640 reconfigurable elements whose size is $2.87 \times 1.42 \times 0.71$ cm$^3$. Each element is composed of two mirror symmetric layers. On each layer, there exists a rectangular copper patch printed on a dielectric substrate, a via hole in the middle, and the ground is placed on the boundary of each layer. As shown in the equivalent circuit in Fig. ~\ref{prototype}, each IOS element comprises two PIN diodes, which connect the patch and the ground. Each PIN diode has two operation states, i.e., ON and OFF, which are controlled by applying bias voltages on the via hole. The dielectric material is PTFE woven glass with permittivity 2.2. Since each IOS element has two PIN diodes, it can have at most four states. However, we only use two states of the IOS for ease of control. The two possible states of each IOS elements are obtained by configuring both PIN diodes in the OFF (state 1) of ON (state 2) states. Table \ref{state} reports the simulated reflection and refraction coefficients of the two states at a working frequency of 3.6 GHz, and for a transmission bandwidth of 24 MHz. These results in Table \ref{state} are obtained with Microwave Studio and the Transient Simulation Package in the CST software, by assuming periodic boundary conditions for the IOS element and normal illumination. The two states of each IOS element correspond to two different and well-defined pairs of the amplitude and phase of the reflection and refraction coefficients. It is worth mentioning that the IOS does not amplify the incident signals since the power (that depends on the square of the amplitudes in Table \ref{state}) of the reflected and refracted signals is always less than 1, i.e., the power of incident signal. For reference, in particular, Table \ref{state} reports the amount of power that is reflected and refracted for each of the two states of the IOS element. The equivalent circuit of each IOS element is also shown in Fig. ~\ref{prototype}. The manufactured IOS prototype does not enable the adaptive control of the amount of power that is reflected and refracted, which is determined by the specific design of the IOS element. This functionality will be added in future implementations of the IOS. 

The IOS is divided into $16$ groups each comprising  $5 \times 8$ elements. For ease of control, the elements within the same group are set to the same state. As shown in Fig. \ref{prototype}, the states of the groups are controlled by an IOS controller, which is implemented by a field-programmable gate array (FPGA). A program is pre-loaded in the FPGA, and the states of the IOS are changed automatically by controlling the FPGA. The FPGA is a Cyclone IV EP4CE10F17C8 platform that is manufactured by Intel Altera corporation. The development board that contains the FPGA and various interfaces is manufactured by ALINX. The hardware description language used for programming the FPGA is Verilog.

	

\vspace{-3mm}
\subsection{Hardware Modules of the Prototype}
The IOS prototype includes the following components.
\begin{itemize}
    \item \emph{Transmitter:} The transmitter (Tx) is implemented by using a universal software radio peripheral (USRP). The type of USRP is LW N210 with an SBX-LW120 RF daughterboard. The USRP implements the functions of RF modulation/demodulation and baseband signal processing by using a GNU Radio software development kit in Python~\cite{OHA-2015}. The output port of the USRP is connected to a ZX60-43-S+ low-noise amplifier (LNA), which amplifies the transmitted signal. A directional double-ridged horn antenna is employed. The antenna part number is LB-880 and is manufactured by A-INFO Corporation.
    
    \item \emph{Receiver:} Similar to the TX, the receiver (Rx) is a USRP, whose input port is connected to an LNA, and directional double-ridged horn antennas are utilized. An external clock (10MHz OCXO) is used to provide a precise clock signal to the Tx and Rx.
    
    \item \emph{Signal synchronizer:} To detect the relative phases and amplitudes of the received signals with respect to the transmitted signals, we use a RIGOL DG4202 signal source generator to obtain the reference signals. The signal source outputs a reference clock signal and a pulse-per-second (PPS) signal to both the Tx and Rx, which are utilized for modulation and demodulation.
    
    \item \emph{Ethernet switch:} The ethernet switch, which connects the Tx, the Rx, and a host computer, enables the acquisition of the transmitted and received signals. The bandwidth of the switch is of the order of 1 GHz.
    
    \item \emph{Data processor:} The data processor is a host computer that controls the Tx and Rx through a software program written in Python. The host computer processes the received signals as well.
\end{itemize}

The different modules of the IOS prototype (in particular the USRP and the FPGA) are coordinated through a software program written in Python on a host computer. Specifically, the host computer configures the USRP through the GNU Radio software development kit. The FPGA, in turn,  transforms the input from the host computer into a control signal for the IOS with the aid of a program written in Verilog.

\vspace{-3mm}
\section{Experimental Evaluation}
\label{sec:experiment}

In this section, we report some experimental results obtained with the manufactured IOS prototype. We first illustrate the experimental setup and then discuss the results.
\vspace{-3mm}
\subsection{Experimental Setup}
In Fig.~\ref{experiment_setup}, the experimental environment and the radiation patterns of the antennas are illustrated. The prototype is deployed in a large and empty room, where the walls are covered with wave absorbing materials, in order to eliminate external interference. A Tx with a single antenna transmits signals to two single-antenna Rx with the aid of the IOS. To show that the IOS can achieve beamforming on both sides of the surface, we deploy a Tx on one side of the IOS, one Rx on the same side of the Tx, and another Rx on the opposite side of the IOS. Both Rx are located at the same distance from the center of the IOS. To clearly observe the beam generated from the IOS, the directional Tx antenna is placed perpendicularly to the IOS, so as to reduce the light-of-sight signals to the Rx on the same side, i.e., Rx 1 in Fig.~\ref{experiment_setup}. The IOS is kept fixed at the center of the room. The distance between the Tx and the IOS is 1.16~m and the distance from the IOS to both Rx is the same and is equal to 0.7~m.

As far as the link budget is concerned, it depends on 1) the transmit power; 2) the antenna gain of the Tx; 3) the channel gain between the Tx and IOS; 4) the gain of the IOS; 5) the channel gain between the IOS and Rx; 6) the antenna gain of the Rx; and 7) the gain of the LNA. In the considered experimental deployment we have the following: the transmit power is 1~dBm; the gains of the Tx and Rx are identical and equal to 10~dB; the channel gain between the Tx and the IOS is -47.76~dB and the channel gain between the IOS and the Rx is -43.53~dB; and the gain of the LNA is 14.3 dB. Finally, the gain of the IOS depends on its configuration according to the amplitude reflection/refraction coefficients reported in Table \ref{state}.

\vspace{-3mm}
\subsection{Experimental Results}
\begin{figure*}[!t]
	\centering
	\includegraphics[width=0.70\textwidth]{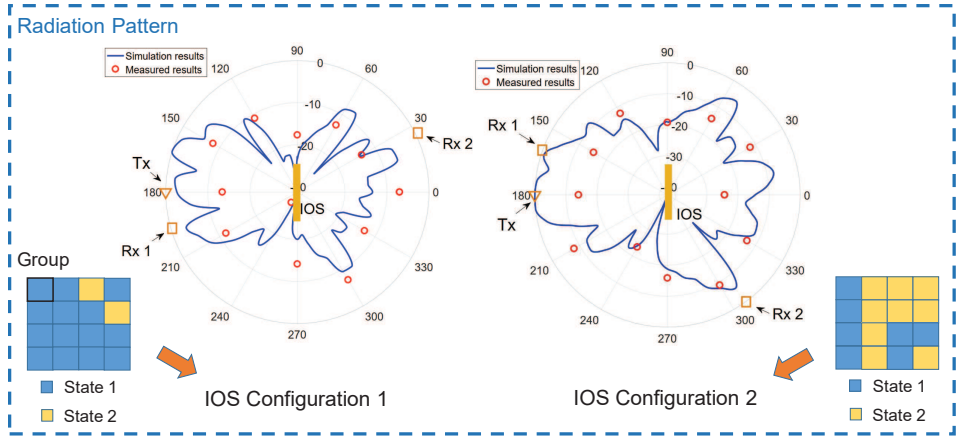}
	\vspace{-2mm}
	\caption{Experimental results: Radiation pattern of the IOS prototype.}
	\vspace{-2mm}
	\label{radiation_pattern_IOS}
\end{figure*}

Fig.~\ref{radiation_pattern_IOS} shows the radiation pattern for two different states of the IOS. In particular, Fig. \ref{radiation_pattern_IOS} compares the radiation patterns obtained through simulations and experimental measurements. The received signal-to-noise ratio (SNR) of Rx 1 and Rx 2 is 14 dB and 8 dB, respectively. This figure shows a good agreement between simulations and measurements, and validates that the manufactured IOS prototype and the proposed beamforming algorithm can generate directional beams towards the intended users located on both sides of the IOS. More specifically, we observe that two different configurations of the IOS (exemplified by the two colormaps) result in two different directional radiation patterns.

\begin{figure*}[!t]
	\centering
	\includegraphics[width=0.70\textwidth]{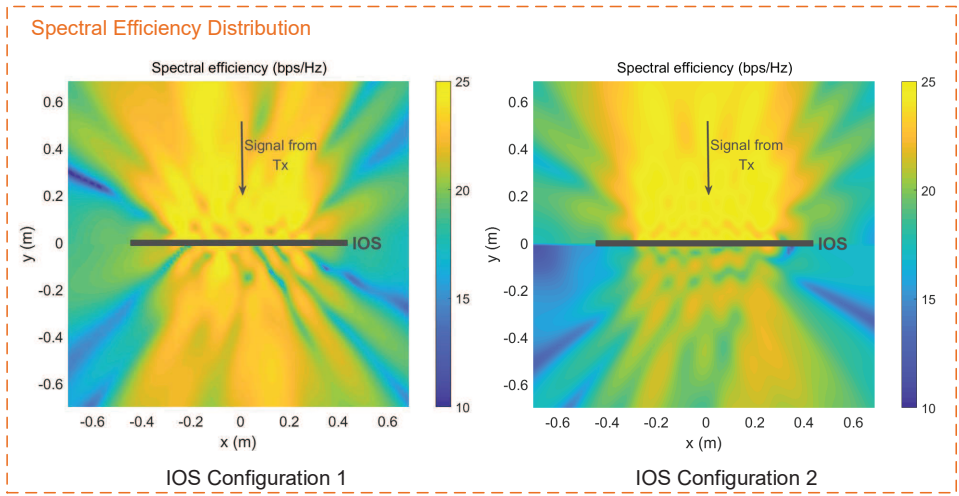}
	\vspace{-2mm}
	\caption{Simulation results: Spectral efficiency distribution.}
	\vspace{-6mm}
	\label{spectral_IOS}
\end{figure*}

Fig. \ref{spectral_IOS} illustrates the spectral efficiency distribution for two different configurations of the IOS.
From this figure, we see that full-dimensional communications can be achieved by the IOS. In addition, we observe that the spectral efficiency distribution depends on the configuration of the IOS elements. This shows that it is necessary to appropriately adjust the states of IOS elements according to the spatial distribution of the users in order to achieve the maximum system capacity.

\vspace{-3mm}
\section{Potential Use Cases and Research Challenges}
\label{sec:applications}
In the previous sections, we have illustrated the potential benefits brought by an IOS and have demonstrated the feasibility of utilizing the IOS for beamforming with the aid of numerical simulations and experimental results. In this section, we briefly present some potential applications of the IOS and future research directions. 
\vspace{-3mm}
\subsection{Coverage Extension}
As illustrated and validated in the previous sections with the aid of experimental results, an IOS has the capability to serve users on both sides of the surface. Therefore, an important application of the IOS is to extend the coverage of cellular networks. By deploying an IOS at the cell edge, for example, the IOS can not only improve the performance of the users within the cell coverage, but can also serve users outside the cell coverage. In addition, the deployment of the IOS brings some unique challenges, which include the following.
\begin{itemize}
    \item \emph{Beamforming design:} To obtain a larger coverage, it is necessary to jointly consider the design of the digital beamforming at the BS and the IOS analog beamforming. However, due to practical constraints, e.g., the quantization of the phase shifts, efficient algorithms need to be developed. Moreover, how to achieve a good performance trade-off on both sides of the IOS is a major research challenge.
    
    \item \emph{Channel acquisition:} The performance achievable by the IOS analog beamforming is highly dependent on the accuracy of the acquired CSI. Since an IOS does not have the capability to digitally process the signals, only the cascaded channels, i.e., the BS-IOS-user channels, can be estimated, whose overhead depends on the number of IOS elements to be estimated \cite{CSGACRMM-2020}. In addition, the channels at both sides of the IOS are correlated, which requires to estimate the CSI jointly.
    
    \item \emph{IOS deployment:} The performance offered by an IOS depends on its relative location with respect to the transmitter and the users. Identifying the optimal deployment of an IOS by taking into account the trade-off between its refractive and reflective functions is an open research issue.   
\end{itemize}
\vspace{-3mm}
\subsection{Interference Cancellation}
An apparent trend in future cellular networks is the ultra-dense deployment of small cells. As a result, the coverage of densely deployed small cells may likely overlap, which enhances the multi-cell interference. In this context, another important application of the IOS in wireless networks is to alleviate the multi-cell interference. This can be achieved by deploying an IOS in the overlapping coverage areas of the small cells. For example, the IOS may be configured to appropriately reflect and refract the signals towards the intended cell-edge users while nulling the signals towards non-intended cell-edge users.

Since multiple small cells may share the same IOS, the coordination among the small cells is a challenging open research issue. In fact, each small cell BS has only access to the CSI of its associated users but the analog beamforming applied by the IOS will affect the users associated to all nearby small cells. Therefore, efficient protocols to coordinate the small cell BSs are necessary. In this context, machine learning may be exploited to effectively tackle the coordination of multiple small cells in complex wireless environments. A similar idea can be applied to device-to-device communications and cognitive radio systems.
\vspace{-3mm}
\subsection{Secure Communications}
Physical layer security is an efficient technology to enable the exchange of confidential messages over a wireless medium in the presence of unauthorized eavesdroppers, without relying on higher-layer encryption. The basic idea is to exploit the inherent randomness of noise and fading channels to limit the amount of information that can be extracted by an unauthorized receiver \cite{ASJA-2014}. With the help of an IOS, the potential of physical layer security can be further enhanced. For example, a user located on one side of the IOS can receive the signals emitted by a transmitter located on the other side of the IOS, but the confidential messages transmitted by this user may not be reflected towards the eavesdroppers that are located on the same side of the user. This can be realized through the optimization of the states of the IOS elements, and, therefore, secure communications can be achieved by capitalizing on the non-reciprocal channel created by the IOS. 

However, IOS-assisted physical layer security poses new research challenges. In the presence of an IOS, the wireless channels become programmable. This requires us to revisit the paradigm of physical later security under the new design constraints that the distribution of the system inputs can be adapted to the states of the system itself. In the context of physical layer security, it is often difficult to have any prior information on the eavesdroppers. Therefore, the optimization of the IOS in order to ensure the desired degree of security is a challenging research issue. In a multi-user system, many users may share the same IOS and the optimization of the cooperation and competition among the users is a major open research problem. 

\vspace{-3mm}
\subsection{Sensing and Localization}
The use of radio-frequency (RF) signals to realize sensing and localization has attracted growing interest due to the low cost and privacy preservation. The idea underlying RF sensing and localization is to exploit specific environmental-dependent features of the wireless signals \cite{JHBLKLYZH-2020}. To achieve a high accuracy, it is important that the signals received at two different locations are as much distinguishable as possible. In this context, the IOS constitutes a promising technology for RF sensing and localization since it can actively customize the propagation channels and enhance their differences. In addition, the full-dimensional communication capability of the IOS can effectively reduce out-of-coverage areas.

The design of IOS-assisted sensing and localization systems brings, however, several challenging problems to solve. One of them is the optimization of the IOS analog beamforming to reduce the sensing and localization errors. Since, in many applications, the signals are sparse in some domains, compressed sensing methods may be used~\cite{JHBLKLYZH-2020}. Machine learning methods may we used as well, in order to classify the received signals based on the presence of objects and the locations of users.

\vspace{-3mm}
\section{Conclusion}
\label{sec:conclusion}
In this article, we have introduced the concept of IOS to realize full-dimensional wireless communications. The working principle of the IOS have been introduced, based on which a hybrid beamforming scheme for IOS-based wireless communications has been discussed. Notably, we have presented the prototype of an IOS-assisted wireless system. Experimental results have substantiated the capability of the IOS to serve users on both sides of the surface and have unveiled that the spectral efficiency is highly related to the configuration of the IOS. Potential applications and open research problems in the context of IOS-assisted wireless communications have been discussed, including coverage extension, interference cancellation, secure communications, as well as sensing and localization.

\end{document}